\documentclass[prd,preprintnumbers,pacs,12pt]{revtex4}
\usepackage{epsfig}
\usepackage{graphicx}

\usepackage{bm}
\usepackage{amsmath}
\usepackage{amssymb}
\usepackage{amscd}
\usepackage{latexsym}

\newcommand{\be}{\begin{equation}}

\newcommand{\ee}{\end{equation}}
\newcommand{\bea}{\begin{eqnarray}}
\newcommand{\eea}{\end{eqnarray}}

\newcommand{\nn}{\nonumber}
\newcommand{\de}{\partial}

\begin{document}

\newcommand{\bra}[1]{\langle #1|}
\newcommand{\ket}[1]{|#1\rangle}
\newcommand{\braket}[2]{\langle #1|#2\rangle}
\newcommand{\tr}{\textrm{Tr}}
\newcommand{\lag}{\mathcal{L}}
\newcommand{\mbf}[1]{\mathbf{#1}}
\newcommand{\desl}{\slashed{\partial}}
\newcommand{\Desl}{\slashed{D}}
\title{A linear moose model with pairs of degenerate gauge boson triplets}

\author{Roberto Casalbuoni, Francesco Coradeschi, Stefania De Curtis and Daniele Dominici}
\affiliation{Department of Physics, University of Florence, and
INFN, 50019 Sesto F., Firenze, Italy}


\begin{abstract}
The possibility  of a strongly interacting
electroweak symmetry breaking sector, as opposed to the weakly
interacting light Higgs of the Standard Model, is not yet ruled out
by experiments. In this paper we make an extensive study of a
deconstructed model (or ``moose'' model) providing an effective
description of such a strong symmetry breaking sector, and show its
compatibility with experimental data for a wide portion of the model
parameter space. The model is a direct generalization of the
previously proposed D-BESS model.

\end{abstract}

\maketitle

\section{Introduction}

Among the problems still left open by the Standard Model (SM) of
particle physics, the understanding of the exact mechanism that
leads to the breaking of the electroweak symmetry at low energies is
of particular importance. Besides the SM basic Higgs mechanism,
still to be verified by experiments, possible alternative solutions
to the problem are offered by extensions of the old technicolor (TC)
theories, where the Higgs boson is realized as a composite state of
strongly interacting fermions.

These theories have recently received a renewed attention starting
from higher dimensional Lagrangians; effective chiral Lagrangians in
four dimensions containing new resonance states can be obtained by
the deconstruction technique
\cite{ArkaniHamed:2001ca,Arkani-Hamed:2001nc,Hill:2000mu,Cheng:2001vd,Abe:2002rj,Falkowski:2002cm,Randall:2002qr,Son:2003et,deBlas:2006fz}
or as holographic versions of 5-dimensional (5D) theories in warped
background \cite{Nomura:2003du,Barbieri:2003pr,Barbieri:2004qk}.

Models have been proposed, working in the framework suggested by the
AdS/CFT correspondence, which assume a $SU(2)_L\times SU(2)_R\times
U(1)_{B-L}$ gauge group in the 5D bulk,
\cite{Csaki:2003dt,Agashe:2003zs,Csaki:2003zu,Cacciapaglia:2004zv,Cacciapaglia:2004rb,Cacciapaglia:2004jz,Contino:2006nn},
or also simpler with a $SU(2)$ in the bulk
\cite{Foadi:2003xa,Hirn:2004ze,Casalbuoni:2004id,Chivukula:2004pk,Georgi:2004iy}.
One of the main challenges of these models is the  value of the $S$
parameter \cite{Peskin:1990zt,Peskin:1991sw} or the related
$\epsilon_3=g^2 S/(16\pi)$
\cite{Altarelli:1990zd,Altarelli:1991fk,Altarelli:1993sz}. Indeed,
the experimental value of $\epsilon_3$ is of the order  of $10^{-3}$
\cite{Barbieri:2004qk}, whereas the value naturally expected in TC
theories is an order of magnitude bigger.

A delocalization of the fermionic fields into the bulk as in
\cite{Cacciapaglia:2004rb,Foadi:2004ps}, realized in the
deconstructed version by allowing  standard fermions to have direct
couplings to all the moose gauge fields as in
\cite{Casalbuoni:2005rs}, leads to direct contributions to the
electroweak parameters that can correct the bad behavior of the
$\epsilon_3$ parameter. The fine tuning which cancels out the
oblique and direct contributions to $\epsilon_3$ in each bulk point,
that is from each internal moose gauge group, corresponds to the so
called \emph{ideal delocalisation} of fermions,
\cite{Casalbuoni:2005rs,Chivukula:2005ji,SekharChivukula:2005cc,Bechi:2006sj,Casalbuoni:2007xn}.
Other solutions to get a suppressed contribution to $\epsilon_3$
have been investigated, like the one suggested by holographic QCD,
assuming that different five dimensional metrics are felt by the
axial and vector states
\cite{Hirn:2005vk,Hong:2006si,Hirn:2006nt,Hirn:2006wg}. However it
has been shown recently that the backgrounds that allow a negative
oblique contribution to $\epsilon_3$ are pathological, since they
require unphysical Higgs profile or higher dimensional operators
\cite{Agashe:2007mc}.

An alternative solution to the $\epsilon_3$ problem was proposed in
\cite{Casalbuoni:1988xm,Casalbuoni:1995yb,Casalbuoni:1995qt} (see
also  \cite{Appelquist:1999dq}). The solution was realized in terms
of an effective TC theory  of non linear $\sigma$-model scalars and
massive gauge fields. The model is a four site model with three
sigma fields. The physical spectrum consists of three massless
scalar fields (the Goldstone bosons  giving masses to the gauge
vector particles) and two triplets of massive vector fields
degenerate in mass and couplings.  This model, named degenerate BESS
(Breaking  Electroweak Symmetry Strongly)
model (D-BESS), has an enhanced custodial symmetry such as to allow
$\epsilon_3=0$ at the lowest order in the electroweak interactions.
This idea has been also recently used in phenomenological analysis
of low scale technicolor theories with vector and axial vector
resonances very close in mass \cite{Eichten:2007sx}.

A generalization of D-BESS was studied in \cite{Casalbuoni:2004id}.
This extended model is a moose model, with a replicated $SU(2)$
gauge symmetry, that maintains the most useful feature of D-BESS,
namely the custodial symmetry which guarantees the vanishing of
$\epsilon_3$.

Two other quantities, $\epsilon_1$ and $\epsilon_2$
\cite{Altarelli:1990zd,Altarelli:1991fk,Altarelli:1993sz} or
equivalently $T$ and $U$ \cite{Peskin:1990zt,Peskin:1991sw}, are
customarily used to parameterize the electroweak precision
observables. More recently, an alternative parametrization was
proposed in terms of seven parameters \cite{Barbieri:2004qk} which
describe in a very general way the effects of so-called
``universal'' extensions of the SM, that is theories whose
deviations from the SM are all contained in the vector boson
self-energies.

In the present paper, we wish to extend the calculation made in
\cite{Casalbuoni:2004id} by deriving the seven parameters of ref.
\cite{Barbieri:2004qk}, and from them the $\epsilon$ parameters, to
the next-to-leading order in  $M_W^2 / M^2$, where $M$ is the
mass scale of the new bosons, and without any expansion in their
gauge couplings. Also, we will calculate, at the same
order, the trilinear gauge boson vertex anomalous contributions due
to the new physics.

In Section \ref{linear} we review the notations and the main
constitutive elements of the model. In Section \ref{efflag} we
derive the low energy effective Lagrangian by eliminating the fields
of the internal moose and show that the model decouples in the limit
$M\to \infty$. In Section \ref{universal} we compute the
effective gauge boson correlation functions, and from them the
parameters of ref. \cite{Barbieri:2004qk} and the $\epsilon$
parameters, to the next-to-leading order in $M_W^2 / M^2$;
we then derive bounds on the model parameter space from experimental
data. In Section \ref{tri} we obtain the effective trilinear gauge
couplings to order $M_W^2/M^2$. Finally, in Section \ref{conclusion}
we present our conclusions.

\section{A linear moose model for the electroweak symmetry breaking}
\label{linear}

Our model is based on the idea of dimensional deconstruction
\cite{ArkaniHamed:2001ca,Arkani-Hamed:2001nc,Hill:2000mu,Cheng:2001vd}
and on the hidden gauge symmetry approach, historically applied both
to strong interactions
\cite{Son:2003et,Bando:1984ej,Bando:1987ym,Bando:1987br} and to
electroweak symmetry breaking
\cite{Casalbuoni:1988xm,Casalbuoni:1995yb,Casalbuoni:1995qt,Casalbuoni:1986vq,Casalbuoni:1985kq}.

Consider $K+1$ non linear $\sigma$-model scalar fields $\Sigma_i$,
${i=1,\cdots ,K+1}$ and $K$ gauge groups, $G_i$, ${i=1,\cdots ,K}$
with global symmetry $G_L\otimes G_R$. Since we are interested in
studying the electroweak symmetry breaking mechanism, we will assume
$G_i \equiv SU(2)$, $G_L\otimes G_R=SU(2)_L\otimes SU(2)_R$. The
transformation properties of the fields are \bea \label{transf}
&&\Sigma_1\to L\Sigma_1 U_1^\dagger,\nn\\
&&\Sigma_i\to U_{i-1}\Sigma_i U_i^\dagger\
,\,\,\,\,\,\,\,i=2,\cdots,K,\nn\\
&&\Sigma_{K+1}\to U_K\Sigma_{K+1} R^\dagger, \eea with $U_i\in G_i$,
$i=1,\cdots,K$; $L\in G_L$, $R\in G_R$; the Lagrangian is given by
\be {\cal L}=\sum_{i=1}^{K+1}f_i^2{\rm Tr}[D_\mu\Sigma_i^\dagger
D^\mu\Sigma_i]-\frac 1 2\sum_{i=1}^K{\rm
Tr}[(\mathbf{F}_{\mu\nu}^i)^2], \label{lagrangian:l} \ee where $f_i$
are $K+1$ free constants (``link'' coupling constants). The
covariant derivatives are defined as follows: \bea
&D_\mu\Sigma_1=\de_\mu\Sigma_1+i\Sigma_1 g_1
\mbf{A}_\mu^1,&\nn\\
&D_\mu\Sigma_i=\de_\mu\Sigma_i-ig_{i-1}\mbf{A}_\mu^{i-1}\Sigma_i+i\Sigma_i
g_i \mbf{A}_\mu^i,&\,\,\,\,\,\,\,i=2,\cdots,K,\nn\\
&D_\mu\Sigma_{K+1}=\de_\mu\Sigma_{K+1}-ig_{K} \mbf{A}_\mu^{K}
\Sigma_{K+1},& \eea as implied by the transformation rules
(\ref{transf}), where $\mbf{A}_\mu^i$ and $g_i$ are the gauge fields
and gauge coupling constants associated to the groups $G_i$,
$i=1,\cdots ,K$. $\mathbf{F}_{\mu\nu}^i$ has the standard definition
\begin{equation}
\mbf{F}_{\mu \nu}^i = \partial_\mu \mbf{A}_\nu^i - \partial_\nu
\mbf{A}_\mu^i + ig_i [\mathbf{A}_\mu^i , \mbf{A}_\nu^i],
\end{equation}
with
\begin{equation}
\mbf{A}_\mu^i = A_\mu^{i,a} \dfrac{\tau^a}{2}.
\end{equation}

Notice that one could introduce an additional field, \be
U=\Sigma_1\Sigma_2\cdots\Sigma_{K+1} \label{chiral}\ee which
transforms just like the usual chiral field of the Higgsless SM:
$U\rightarrow LUR^\dagger$. The field $U$ is an invariant under the
$G_i$ transformations (which are then effectively "hidden" to $U$).

As shown in \cite{Casalbuoni:2004id}, in this model, due to the
presence of a custodial $SU(2)$ symmetry, we get no corrections to
$\epsilon_{1,2}$. Also, if we put one (and only one) of the link
coupling constants $f_i$ equal to zero, we effectively enlarge the
global symmetry to $(SU(2) \otimes SU(2))^{K+1}$, getting
$\epsilon_3 = 0$ too. We want to explore this particular case; also,
for simplicity, we impose an extra left-right symmetry of the moose
which identifies the two ends:
\begin{gather}
f_i \equiv f_{K+2-i},\\
g_i \equiv g_{K+1-i}.
\end{gather}
The reflection symmetry, together with the condition that just one
of the link coupling constants must vanish, implies that the number
of moose links has to be odd (and hence the number of gauge fields
has to be even), and that the vanishing link has to be the central
one. So we will consider:
\begin{gather}
K = 2N,\\
f_{N+1} = 0.
\end{gather}

It is instructive to count out the number of degrees of freedom.
Before cutting the central link, we had $(2N+1)$ matrices of scalar
fields, for $3(2N+1)$ degrees of freedom, and $3(2N)$ massless
vector fields, for $6(2N)$ degrees of freedom; of these, only 3
scalar fields are physical, the others disappearing (in the unitary
gauge) to give mass to all the gauge bosons via the Higgs mechanism.
After the cutting, we only get $3(2N)$ scalar degrees of freedom to
start with, so that no one survives the Higgsing of the gauge
bosons.

It will be useful for further considerations to look at the form of
the gauge boson mass matrix, which can be obtained by putting
$\Sigma_i=I$ in eq. (\ref{lagrangian:l}). We find \be
\lag_{\rm{mass}}=\sum_{i=1,\ i\neq N+1}^{2N+1}f_i^2
\textrm{Tr}[(g_{i-1}\mbf{A}_\mu^{i-1} -g_i \mbf{A}_\mu^i)^2]\equiv
\sum_{i,j=1}^{2N}(M_2)_{ij} \textrm{Tr}[\mbf{A}_\mu^i \mbf{A}^{j
\;\! \mu}], \label{lmass} \ee with \be
\begin{split}
(M_2)_{ij}= & g_i^2(f_i^2+f_{i+1}^2)\delta_{i,j} - g_i
g_{i+1}f_{i+1}^2
\delta_{i,j-1}-g_j g_{j+1}f_{j+1}^2  \delta_{i,j+1},\\
& i,j = 1, \cdots 2N, \quad g_0=g_{2N+1}=0\,. \label{m2}
\end{split}
\ee Thanks to the condition $f_{N+1} = 0$ and to the reflection
symmetry, the matrix $M_2$ is block diagonal with two degenerate
blocks. Each block can be independently diagonalized through an
orthogonal transformation $S$. By calling  $\mbf{\tilde A}_\mu^i$,
$i=1,\cdots,N$ the mass eigenstates, and $m^2_n$ the squared mass
eigenvalues, we have 
\be \mbf{A}_\mu^i=\sum_{n=1}^N S^i_n
\mbf{\tilde A}_\mu^n,\ee 
with
\be
S^i_m(M_2)_{ij}S^j_n=m^2_n\delta_{m,n},\ee 
and an analogous result
holds for $\mbf{A}^i, \ i=N+1, \cdots 2N$. We will assume $m_n\neq
0, \ n=1, \cdots, 2N$, otherwise the model describes an unphysical
situation.

\section{Calculation of the effective Lagrangian}
\label{efflag} We will now switch on the electroweak interactions by
gauging the $SU(2)_L\otimes U(1)_Y$ subgroup of the global
$G_L\otimes G_R$. We will include in the model only standard
fermions coupled to $SU(2)_L\otimes U(1)_Y$.  Then, considering the
limit of heavy mass for the extra gauge bosons, we will integrate
them out in order to obtain an effective description in terms of the
electroweak and the fermion fields only.

Note that just promoting part of the global symmetry to a gauge
symmetry is not enough to describe realistic $W$ and $Z$ bosons: we
also need to provide suitable mass terms for three out of four of
the newly added gauge fields. In the model as it is, however, there
is not any scalar degree of freedom left to trigger a Higgs
mechanism, as we have seen. A natural way out is to add to the
Lagrangian an additional term containing the chiral field $U$, which
obeys the transformation rule $U \to LUR^\dag$:
\begin{equation}
\lag_U =  f_0^2 \textrm{Tr} [D_\mu U^\dagger D^\mu U],
\label{lagU}
\end{equation}
with
\begin{equation}
\begin{split}
D_\mu U & = \partial_\mu U -i \tilde{g} \mathbf{\tilde{W}_\mu} U + i U \tilde{g}' \mathbf{\tilde{Y}_\mu},\\
\mathbf{\tilde{W}_\mu} & = \tilde{W}_\mu^a \dfrac{ \tau^a}{2} \quad
\mathbf{\tilde{Y}_\mu} = \tilde{Y}_\mu \dfrac{\tau^3}{2}.
\end{split}
\end{equation}

The $U$ field gives us the additional three degrees of freedom we
need and provides a SM-like symmetry breaking term for the gauge
bosons $\tilde{W}$ and $\tilde{Y}$ associated to $SU(2)_L \otimes
U(1)_Y$.

Summing up, the Lagrangian of the bosonic sector of the generalized
D-BESS (GD-BESS) model is
\begin{equation}
\begin{split}
\label{eq:1}
\mathcal{L} & =\sum_{i=1, i \neq N+1}^{2N+1} f_i^2 \textrm{Tr}[D_\mu\Sigma_i^\dag D^\mu \Sigma_i]  + f_0^2 \textrm{Tr} [D_\mu U^\dagger D^\mu U]\\
& - \frac{1}{2}\textrm{Tr}[(\mathbf{F}_{\mu \nu}^{\tilde{W}})^2]-
\frac{1}{2}\textrm{Tr}[(\mathbf{F}_{\mu \nu}^{\tilde{Y}})^2]-
\frac{1}{2}\sum_{i=1}^{2N}\textrm{Tr}[(\mbf{F}_{\mu \nu}^i)^2]
\end{split}
\end{equation}
where:
\begin{equation}
\label{fields}
\begin{array}{l}
\mathbf{F}_{\mu \nu}^{\tilde{W}} = \partial_\mu \mathbf{\tilde{W}_\nu} - \partial_\nu \mathbf{\tilde{W}_\mu} + i\tilde{g} [\mathbf{\tilde{W}_\mu , \tilde{W}_\nu}],\\
\mathbf{F}_{\mu \nu}^{\tilde{Y}} = \partial_\mu \mathbf{\tilde{Y}_\nu} - \partial_\nu \mathbf{\tilde{Y}_\mu}\\
\end{array}
\end{equation}
and the covariant derivatives of $\Sigma_1$ and $\Sigma_{2N+1}$ are
modified as follows:
\begin{equation}
\begin{split}
D_\mu\Sigma_1 & = \de_\mu\Sigma_1-i \tilde{g}\mbf{\tilde{W}}_\mu \Sigma_1 + i\Sigma_1 g_1 \mbf{A}_\mu^1,\\
D_\mu\Sigma_{2N+1} & = \de_\mu\Sigma_{2N+1} - i
g_{2N}\mbf{A}_\mu^{2N} \Sigma_{2N+1} + i \tilde{g}' \Sigma_{2N+1}
\mbf{\tilde{Y}}_\mu
\end{split}
\end{equation}
due to the gauging of $SU(2)_L\otimes U(1)_Y$.

The fermion interactions will be given by SM-like terms:
\begin{equation}
\label{ferm}
\begin{split}
\lag_{fermion} = & - \tilde{g} \, \overline{\psi} \gamma^\mu
\frac{(1-\gamma^5)}{2} \frac{\tau^a}{2} \psi \,\tilde{W}_\mu^a -
\tilde{g}' \, \overline{\psi}
\gamma^\mu\frac{(1-\gamma^5)}{2}\frac{B-L}{2} \psi \,\tilde{Y}_\mu\\
& - \tilde{g}' \, \overline{\psi} \gamma^\mu \frac{(1+\gamma^5)}{2}
\left(\frac{B-L}{2}+\frac{\tau^3}{2} \right) \psi \,\tilde{Y}_\mu
\end{split}
\end{equation}
where $\psi$ is a generic fermion doublet, and $B$, $L$ are the
barion and lepton numbers respectively. In this way, the new gauge
bosons are coupled to the fermions only through their mixing with
the SM ones.

By expanding eq. (\ref{eq:1}) in the unitary gauge $\Sigma_i = I, \
\forall i$, and separating the kinetic term contribution from that
of the terms containing the link coupling constants, we get
\begin{eqnarray}
\label{lageffsep1}
\mathcal{L}_{kin} & = & - \frac{1}{4} \sum_{i=0}^{2N+1}(A_{\mu \nu}^{i,a}  - g_i \epsilon^{abc} A_\mu^{i,b} A_\nu ^{i,c})^2, \\
\label{lageffsep2} \mathcal{L}_{link} & = &
\sum_{\genfrac{.}{.}{0pt}{}{i=1}{(i \neq N+1)}}^{2N+1}
\frac{f_i^2}{2} (g_{i-1}A_\mu^{i-1,a} - g_i A_\mu^{i,a})^2 +
\frac{f_0^2}{2} (\tilde{g} \tilde{W}_\mu^a - \tilde{g}'
\tilde{Y}_\mu^a)^2,
\end{eqnarray}
where we have made the identifications:
\begin{equation}
A_\mu^{0,a}=\tilde{W}_\mu^{a}, \quad A_\mu^{2N+1,3}=\tilde{Y}_\mu,
\quad A_\mu^{2N+1,1}=A_\mu^{2N+1,2}=0, \quad g_0 = \tilde{g}, \quad
g_{2N+1} = \tilde{g}'
\end{equation}
and defined:
\begin{equation}
A_{\mu \nu}^{i,a} = \de_\mu A^{i,a}_\nu - \de_\nu A_\mu^{i,a}, \quad
i=0, \cdots, 2N+1.
\end{equation}

\begin{figure}[h] \centerline{
\epsfxsize=12cm\epsfbox{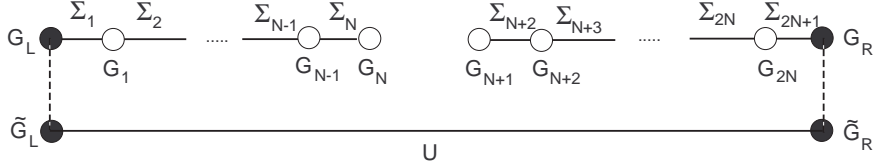} } \caption {{  Moose
diagram for GD-BESS; it consists of a linear moose with the central
link cut and a nonlocal connection (the $U$ field) between the two
end points of the moose. $\tilde{G}_L \otimes \tilde{G}_R$ is the
global symmetry group after the gauging of the electroweak
subgroup.} \label{fig:1} }
\end{figure}

The model field content is summarized in Fig. \ref{fig:1}. For $N=1$
the model reduces to the D-BESS model
\cite{Casalbuoni:1995yb,Casalbuoni:1995qt}.

From the Lagrangian (\ref{eq:1}), we can derive the classical
equations of motion for the $\mbf{A}_\mu^i$ fields:
\begin{equation}
\label{moto}
\begin{split}
 \partial_\mu \mathbf{F}^{i \, \mu \nu} & = ig_i[\mbf{A}_\mu^i,\mathbf{F}^{i \, \nu \mu}] + g_i[f_i^2(g_{i-1}\mathbf{A}^{i-1 \, \nu} - g_i \mathbf{A}^{i \, \nu}) \\
& -f_{i+1}^2(g_i \mathbf{A}^{i\, \nu} - g_{i+1} \mathbf{A}^{i+1 \,
\nu})], \quad i=1, \cdots, 2N
\end{split}
\end{equation}
where again we have identified
\begin{equation}
\mbf{A}^0_\mu = \mbf{\tilde{W}}_\mu, \quad \mbf{A}_\mu^{2N+1} =
\mbf{\tilde{Y}}_\mu.
\end{equation}

If this model is to be consistent with the existing experimental
data, the masses of the $A^{i,a}$ fields must be significantly
larger than those of the SM gauge bosons. Let's call $M$ the common
mass scale of the heavy gauge bosons. Here we will be concerned only
about the low energy predictions of the model, in which the new
particles are not directly produced, but rather manifest themselves
only by small modifications of the SM gauge boson propagators. So we
will use an effective Lagrangian approach and work in the limit $p^2
\ll M^2$, where $p$ represents the typical momentum scale of the
processes we wish our effective theory to describe. Since the mass
spectrum cannot be determined analytically in the general case, we
do not have an exact expression for $M$ but we can give an estimate
of it by looking at the mass matrix (\ref{m2}). We see that all the
terms in (\ref{m2}) are a sum of contributions which are
proportional to $f_i^2 g_j^2$ for some $i,j$; so we can assume that
the typical mass scale will be of order $f_i g_j$ (we will get a
more explicit estimate for the mass scale in the following). This
assumption can actually be checked in the simplest case $N=1$, as
shown by the direct analysis made in ref. \cite{Casalbuoni:1995yb}.
In this case, we have four massive eigenstates; their masses are, to
lowest order: $\tilde{M}_W \simeq \tilde{g} f_0$, $\tilde{M}_Z
\simeq \tilde{g}' f_0$ for the two lightest eigenstates, which can
be identified with the SM gauge bosons, and $M \simeq g_1 f_1$ for
the two heaviest states, which are degenerate.

As a consequence, the limit we will study is
\begin{equation}
p^2 \ll f_i^2 g_j^2, \ \ \ \
i=1,\cdots,N,N+2,\cdots,2N+1;~~~j=1,\cdots,2N.
\end{equation}
If we now rewrite eq. (\ref{moto}) as
\begin{equation}
\label{moto2}
\partial_\mu g_i \mathbf{F}^{i \, \mu \nu} + i[g_i \mbf{A}_\mu^i, g_i\mathbf{F}^{i \, \mu \nu}] =
g_i^2 [f_i^2(g_{i-1}\mbf{A}^{i-1 \, \nu} - g_i \mbf{A}^{i \, \nu}) -
f_{i+1}^2 (g_i \mbf{A}^{i \, \nu} - g_{i+1} \mbf{A}^{i+1 \, \nu})],
\end{equation}
we can see that the quantities on the left-hand side are of higher
order with respect to those on the right-hand side. Keeping only
leading order terms, the equations reduce to
\begin{equation}
f_i^2(g_{i-1} \mathbf{A}_\mu^{i-1} - g_i \mbf{A}_\mu^i) - f_{i+1}^2(
g_i \mbf{A}_\mu^i - g_{i+1} \mathbf{A}_\mu^{i+1}) = 0,
\end{equation}
which imply
\begin{equation}
\label{leadW} g_N\mathbf{A}_\mu^N = g_{N-1} \mathbf{A}_\mu^{N-1} =
\ldots =  g_1 \mathbf{A}_\mu^1 = \tilde{g} \mathbf{\tilde{W}}_\mu
\end{equation}
and
\begin{equation}
\label{leadY} g_{N+1} \mathbf{A}_\mu^{N+1} = g_{N+2}
\mathbf{A}_\mu^{N+2} = \ldots = g_{2N} \mathbf{A}_\mu^{2N}=
\tilde{g}' \mathbf{\tilde{Y}}_\mu.
\end{equation}
Substituting this leading order expressions in the unitary gauge
Lagrangian given in eqs. (\ref{lageffsep1}),(\ref{lageffsep2}),
limiting us for the moment to the bilinear terms, we get:
\begin{equation}
\label{eq:2}
\begin{split}
\mathcal{L}_{eff}^2 & = -\frac{1}{2} (\frac{1}{\tilde{g}^2}+\frac{1}{\overline{G}^2})\tilde{g}^2 \tilde{W}_{\mu \nu}^+ \tilde{W}^{- \, \mu \nu} -\frac{1}{4} (\frac{1}{\tilde{g}^2}+\frac{1}{\overline{G}^2})\tilde{g}^2 \tilde{W}_{\mu \nu}^3 \tilde{W}^{3 \, \mu \nu}- \frac{1}{4} (\frac{1}{\tilde{g'}^{2}}+\frac{1}{\overline{G}^2})\tilde{g'}^{2} \tilde{Y}_{\mu \nu} \tilde{Y}^{\mu \nu}\\
& + \tilde{g}^2 f_0^2 \tilde{W}_\mu^+ \tilde{W}^{- \, \mu} +
\frac{\tilde{g}^2 f_0^2}{2}\tilde{W}_\mu^3 \tilde{W}^{3 \, \mu} +
\frac{\tilde{g}'^2 f_0^2}{2} \tilde{Y}_\mu \tilde{Y}^\mu - f_0^2
\tilde{g}\tilde{g}' \tilde{W}_\mu^3 \tilde{Y}^\mu,
\end{split}
\end{equation}
where
\begin{equation}
\frac{1}{\overline{G}^2} = \sum_{k=1}^{N} \frac{1}{g_k^2} =
\sum_{k=N+1}^{2N} \frac{1}{g_k^2} \label{Gi}
\end{equation}
and we have introduced the charged gauge fields $\tilde{W}^\pm=
\dfrac{1}{\sqrt{2}}(\tilde{W}^1 \mp i \tilde{W}^2)$. This expression
exactly reproduces the SM electroweak gauge Lagrangian, provided we
rescale the fields $\tilde{W}$, $\tilde{Y}$: $\tilde{g} \tilde{W}
\to g \tilde{W}$, $\tilde{g}' \tilde{Y} \to g' \tilde{Y}$, and
identify
\begin{equation}
\label{g} \frac{1}{g^2} =
(\frac{1}{\tilde{g}^2}+\frac{1}{\overline{G}^2}), \quad
\frac{1}{g'^{2}} =
(\frac{1}{\tilde{g'}^{2}}+\frac{1}{\overline{G}^2}), \quad f_0^2 =
\frac{v^2}{4} \equiv \frac{(\sqrt{2} G_F)^{-1}}{4},
\end{equation}
with
\begin{equation}
f_0^2g^2 \simeq M_W^2, \quad f_0^2 (g^2+g'^{2}) \simeq M_Z^2
\end{equation}
in the limit $M \to \infty$. This means that any deviation from the
SM at low energy will be suppressed at least by a factor
$\frac{p^2}{M^2}$.

We can now get the next-to-leading order expression for the
$\mathbf{A}^i$ iteratively, by substituting the leading order
solutions (\ref{leadW})-(\ref{leadY}) in the left-hand side of eq.
(\ref{moto2}). We get
\begin{gather}
\label{Ai}
g_i \mbf{A}_\nu^i = g \mathbf{\tilde{W}}_\nu - c_i \mbf{K}_\nu, \ \ i=1, \ldots, N; \\
\label{Ai2} g_i \mbf{A}_\nu^i = g' \mathbf{\tilde{Y}}_\nu - c_i
\mbf{H}_\nu , \ \ i = N+1, \ldots, 2N;
\end{gather}
where we have introduced:
\begin{equation}
\begin{split}
\label{ci}
\mbf{K}_\nu = & \; g \partial^\mu \mathbf{F}^{\tilde{W}}_{\mu \nu} + ig^2 [\mathbf{\tilde{W}}^\mu, \mathbf{F}^{\tilde{W}}_{\mu \nu}], \quad \mbf{H}_\nu = \, g' \partial^\mu \mathbf{F}^{\tilde{Y}}_{\mu \nu},\\
c_i = \displaystyle{\sum_{j=1}^i} & \dfrac{1}{f_j^2} \sum_{k=j}^N
\frac{1}{g_k^2} = c_{N+i} = \displaystyle{\sum_{j=N+i}^{2N+1}}
\dfrac{1}{f_j^2} \sum_{k=N+1}^{j}\frac{1}{g_k^2}, \quad i=1, \ldots,
N.
\end{split}
\end{equation}
Notice that the $c_i$ are positive definite and of order
$O(\frac{1}{M^2})$, and the reflection symmetry implies $c_i =
c_{2N+1- i}$.

Let us make the substitution for the $\mathbf{A}^i$. Limiting us to
the quadratic part of the Lagrangian and  using eqs.
(\ref{Ai})-(\ref{Ai2}), we get:
\begin{equation}
\label{eq:3}
\begin{split}
& \mathcal{L}_{eff}^2 = -\frac{1}{2} \tilde{W}_{\mu \nu}^+ \tilde{W}^{- \, \mu \nu} -\frac{1}{4}\tilde{W}_{\mu \nu}^3 \tilde{W}^{3 \, \mu \nu}- \frac{1}{4}\tilde{Y}_{\mu \nu} \tilde{Y}^{\mu \nu}\\
+ & \frac{v^2 g^2}{4} \tilde{W}_\mu^+ \tilde{W}^{- \, \mu} + \frac{v^2 g^2}{8}\tilde{W}_\mu^3 \tilde{W}^{3 \, \mu} + \frac{v^2 g'^{2}}{8} \tilde{Y}_\mu \tilde{Y}^\mu - \frac{v^2 g g'}{4} \tilde{W}_\mu^3 \tilde{Y}^\mu \\
+ & \dfrac{1}{4 \overline{G}^2} \frac{1}{\overline{M}^2} \left( 2g^2
\tilde{W}_{\mu \nu}^+ \square \tilde{W}^{- \, \mu \nu} + g^2
\tilde{W}_{\mu \nu}^3 \square \tilde{W}^{3 \, \mu \nu} + g'^{2}
\tilde{Y}_{\mu \nu} \square \tilde{Y}^{\mu \nu} \right);
\end{split}
\end{equation}
where
\begin{gather}
\frac{1}{\overline{M}^2}= C\overline{G}^2, \quad C \equiv  \sum_{i =
1}^N \dfrac{{c_i}}{g_i^2} \equiv  \sum_{i = N+1}^{2N}
\dfrac{{c_i}}{g_i^2}.
\end{gather}
$\overline{M}$ can be used as an explicit estimate for the scale
$M$ (from the definition of the $c_i$ in (\ref{ci}), we 
see that $\overline{M}$ is indeed of order $f_i g_j$).


\section{Effective gauge boson correlation functions and $\epsilon$ parameters}
\label{universal}

From eq. (\ref{eq:3}), it is straightforward to calculate the
correlators for the fields $\tilde{W}$ and $\tilde{Y}$. Up to the
fourth power of the momentum, we get
\begin{equation}
\label{corr}
\begin{split}
\Pi_{+-}(p^2) & = - \frac{1}{g^2} \left(g^2 \frac{v^2}{4} - p^2 - p^4 \frac{g^2}{\overline{M}^2 \overline{G}^2} \right) \\
\Pi_{33}(p^2) & = -  \frac{1}{g^2} \left(g^2 \frac{v^2}{4} - p^2 - p^4 \frac{g^2}{\overline{M}^2 \overline{G}^2} \right) \\
\Pi_{YY}(p^2) & = -\frac{1}{g'^{2}} \left(g'^{2} \frac{v^2}{4} - p^2 - p^4 \frac{g'^{2}}{\overline{M}^2 \overline{G}^2} \right) \\
\Pi_{3Y}(p^2) & = \frac{v^2}{4}.
\end{split}
\end{equation}
It is immediate to verify that, as it should,
\begin{equation}
\frac{1}{g^2} = \Pi'_{+-}(0), \quad \frac{1}{g'^{2}} = \Pi'_{YY}(0),
\quad v^2 = -4 \Pi_{+-}(0),
\end{equation}
where the derivatives of the $\Pi$ are taken with respect to $p^2$.

Following ref. \cite{Barbieri:2004qk}, one can consider seven form
factors, encoding the corrections of new physics to the electroweak
precision observables:
\begin{equation}
\label{seven}
\begin{split}
\hat{S} = g^2 \Pi_{3Y}' (0), \quad & \hat{T} = g^2 M_W^2 (\Pi_{33} (0) - \Pi_{+-} (0)) \\
\hat{U} = -g^2 (\Pi'_{33} (0) - \Pi'_{+-} (0)), \quad & V = \frac{1}{2} g^2M_W^2 (\Pi''_{33} (0) - \Pi''_{+-} (0)) \\
X = \frac{1}{2} g g' M_W^2 \Pi''_{3Y} (0), \quad Y = &
\frac{1}{2}g'^{2} M_W^2 \Pi''_{YY} (0), \quad W = g^2 M_W^2
\Pi''_{33} (0).
\end{split}
\end{equation}
Notice that the analysis of ref. \cite{Barbieri:2004qk} only applies
to ``universal'' theories; the GD-BESS model belongs to this class
since the couplings with the fermions are of the standard form. In
our model,  from the equality of $\Pi_{+-}$ and $\Pi_{33}$
and the expression for $\Pi_{3Y}$ in eq. (\ref{corr}) it follows that $\hat{S}
= \hat{T} = \hat{U} = V = X = 0$; so we have only two non vanishing
form factors, namely $W$ and $Y$:
\begin{equation}
W = \frac{g^2 M_W^2}{\overline{M}^2 \overline{G}^2}, \quad Y =
\frac{g'^{2}M_W^2}{\overline{M}^2 \overline{G}^2}.
\end{equation}
We can now compare the GD-BESS model predictions to experimental
results. For this purpose, it is convenient to consider the
$\epsilon$ parameters, since they are better constrained by the data
and more widely used in the literature. From the definition of the
$\epsilon$ in terms of the form factors \cite{Barbieri:2004qk}, we
get, as contributions from new physics,
\begin{equation}
\begin{array}{l}
\epsilon_1 = \hat{T} - W + 2\dfrac{ s_{\theta}} {c_{\theta}}X-\dfrac{s_{\theta}^2}{c_{\theta}^2}Y,\\
\epsilon_2 = \hat{U} - W + 2\dfrac{ s_{\theta}} {c_{\theta}}X - V, \\
\epsilon_3 = \hat{S} - W + \dfrac{1}{s_{\theta} c_{\theta}}X - Y,\\
\end{array}
\end{equation}

where $\tan(\theta) = g'/g$. For the GD-BESS model we find:
\begin{equation}
\label{epsilon} \epsilon_1 = - \frac{(c_{\theta}^4 +
s_{\theta}^4)}{c_{\theta}^2} \overline{X}, \quad \epsilon_2 = -
c_{\theta}^2 \overline{X}, \quad \epsilon_3 = - \overline{X},
\end{equation}
with $\overline{X}$ given by
\begin{equation}
\label{X} \overline{X} = \frac{M_Z^2}{\overline{M}^2}
\left(\frac{g}{\overline{G}}\right)^2.
\end{equation}

As we can see, after the gauging of the electroweak interactions,
the new physics contribution to the $\epsilon$ parameters is no
longer equal to zero, but the leading non vanishing order of the
correction is $O(M_Z^2/\overline{M}^2)$. This contribution can be
understood as follows: the weak interactions explicitly break the
custodial $(SU(2) \otimes SU(2))^{2N+1}$, which protects $\epsilon_3
= 0$, down to the standard $SU(2)_L \otimes U(1)_Y$. The case with
no weak interactions can be re-obtained in the limit
$M_Z^2/\overline{M}^2 \to 0$, which represents a zeroth-order
approximation of the model. More generally, all of the SM
electroweak sector will be modified by $O(M_Z^2/\overline{M}^2)$
contributions due to the new physics. As an explicit example of
this, in the following Section \ref{tri} we will calculate the
effective contribution to the trilinear gauge boson couplings.

\begin{figure}[h] \centerline{
\epsfxsize=10cm\epsfbox{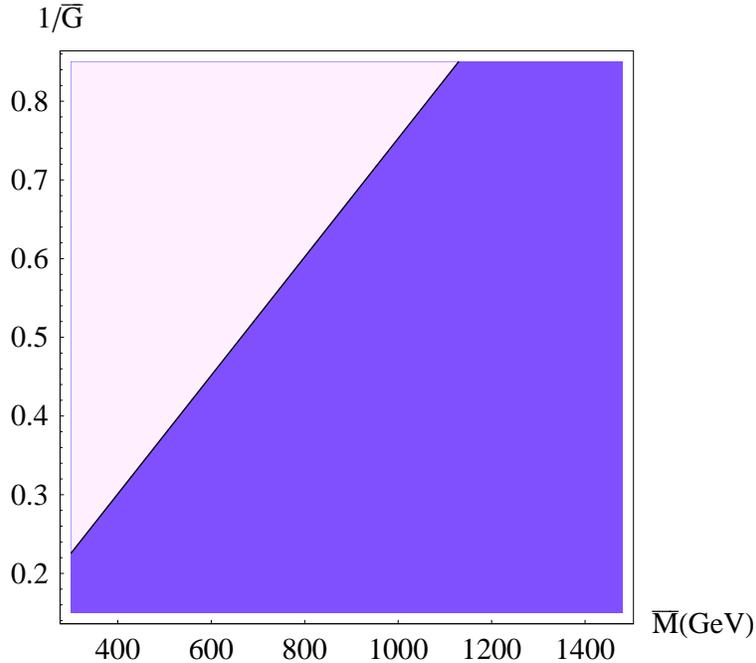} } \caption {{ $95 \%$ C.L.
allowed region (the darker one), in the parameter space
($\overline{M}, 1/\overline{G}$) by  comparison of GD-BESS model
predictions to electroweak precision parameters $\epsilon_1$,
$\epsilon_2$ and $\epsilon_3$. Predictions include radiative
corrections as in the SM with $m_t = 170.9$ GeV and $m_H = 1000$
GeV.} \label{fig:4}  }
\end{figure}

The new physics contribution to the $\epsilon$ parameters can be
tested against experimental data. In Figure \ref{fig:4} we show a
$\chi^2$ contour plot in the plane $(\overline{M},
\dfrac{1}{\overline{G}})$ at $95 \%$ C.L.; the contour is obtained
by considering the following experimental values for the $\epsilon$
parameters:
\begin{equation}
\left.\begin{array}{l}\epsilon_1 = (+5.0 \pm 1.1) 10^{-3} \\
\epsilon_2 = (-8.8 \pm 1.2) 10^{-3}\\ \epsilon_3 = (+4.8 \pm 1.0)
10^{-3} \end{array}\right. \ \textrm{with correlation matrix} \
\left(\begin{array}{ccc}1 & 0.66 & 0.88 \\ 0.66 & 1 & 0.46 \\0.88 &
0.46 & 1\end{array}\right);
\end{equation}
and adding to the present model contributions the SM values:
\begin{equation}
\epsilon_1 =  3.4 \ 10^{-3}, \quad \epsilon_2 =  - 6.5 \ 10^{-3},
\quad \epsilon_3 =  6.7 \ 10^{-3},
\end{equation}
given for $m_t = 170.9$ GeV and assuming an effective $m_H = 1000$
GeV (experimental data are taken from \cite{Barbieri:2004qk} for the
$\epsilon$ parameters and from the Tevatron EWWG web site for the
top mass, while the SM radiative corrections are obtained as a
linear interpolation from the values listed in
\cite{Altarelli:2000ma}). Notice that a relatively low scale
$\overline{M}$ is still allowed by present data. This is due to the
double suppression factor present in $\overline{X}$, eq. (\ref{X}).
Notice also that the GD-BESS model in the limit $\overline{M} \to
\infty$ reproduces the SM to all orders in $\dfrac{g}{\overline{G}}$
(decoupling).

The result can be made much more explicit in the simplest case $f_i
\equiv f_c$ e $g_i \equiv g_c \ \forall i$. Recalling that
\begin{equation*}
\frac{1}{\overline{M}^2}= C\overline{G}^2 = \sum_{i = 1}^N
\frac{{c_i}}{{g_i^2}} \left( \sum_{j=1}^{N} \frac{1}{g_j^2}
\right)^{-1}
\end{equation*}
we have, in this case:
\begin{equation}
\overline{X} = \frac{N(N+1)(2N+1)}{6f_c^2 g_c^2} \left(\frac{g}{g_c}
\right)^2 M_Z^2\,;
\end{equation}
and, from $\overline{X}$, we immediately derive the $\epsilon$
parameters from eq. (\ref{epsilon}). We can verify that with the
substitutions $f_c^2 \to 2 a_2 \frac{v^2}{4}$ and $g_c^2 \to
\frac{g''^2}{2}$, with $N = 1$, eqs. (\ref{epsilon}), (\ref{X})
coincide with the D-BESS result
\cite{Casalbuoni:1995yb,Casalbuoni:1995qt}. Indeed, in this case we
find that
\begin{equation}
\overline{X} = \frac{1}{2a_2 \dfrac{v^2}{4} \dfrac{g''^2}{2}}
\left(\frac{g}{g''} \right)^2 M_Z^2 = 2 \frac{M_Z^2}{M_{BESS}^2}
\left(\frac{g}{g''} \right)^2,
\end{equation}
where $M_{BESS} = \sqrt{a_2} \dfrac{v}{2} {g''}$
 coincides with ${\overline M}$ for $N=1$ and,  as shown in
\cite{Casalbuoni:1995yb,Casalbuoni:1995qt}, represents the mass of the two degenerate new resonances.
Therefore for $N=1$ the limit
shown in Fig. \ref{fig:4} can be interpreted as a bound on the degenerate masses  of the new gauge vectors. 

In general  in order to get limitations from the electroweak precision data
on the mass spectrum,  one needs to perform
the  mass diagonalization  which depends on the specific value of $N$
and also on the
particular choices of $g_i$ and $f_i$. However, for any $N$, $\overline M$
gives the typical mass scale of the lowest resonance. For example,
for $g_i=g_c$, $f_i=f_c$, using the result for the spectrum given
in \cite{Foadi:2003xa}, neglecting the electroweak interactions, the relation between the lightest charged resonance mass and $\overline M$ is 
\be
M^{(1)}=2 \sin{\big(\frac \pi {2 (N+1)}\big)} \sqrt{\frac{(N+1)(2 N+1)}{6}}{\overline M}\sim 1.6-1.8 \,{\overline M}\,~~~~{\rm for}\, N>>
1
\ee

Let us conclude this Section with a comment on the partial
wave unitarity violation. A special feature of GD-BESS model is that
the unitarity bound is completely determined by the $U$ term given in eq. (\ref{lagU}). In fact one can verify explicilty that the scattering amplitudes for the
longitudinal electroweak vector bosons (using the equivalence theorem) are equal
to the ones obtained for the Higgsless SM (for $N=1$ see \cite{Casalbuoni:1997bv}) and that all the amplitudes for the longitudinal $A_i$'s can be always arranged to have a higher unitarity bound \cite{Accomando:2006ga}. Therefore
this model is expected to become strongly interacting around a scale $4\sqrt{\pi}v\sim$ 1.7 TeV. One possible way to unitarize the GD-BESS model is to include
also scalars associated to the $\Sigma_i$ fields on each site and to
the $U$ field. In the simple case of $N=1$ we have shown that the
resulting theory is renormalizable (and unitary) and  decoupling holds
\cite{Casalbuoni:1996wa}. The generalization to generic $N$ is under study.


\section{Trilinear couplings to the next-to-leading order} \label{tri}
We will now calculate, still to the next-to-leading order in the weak
interactions, the contributions of GD-BESS model to the SM trilinear
couplings. These can be read out of the effective trilinear
Lagrangian, which is again obtained by substituting eqs. (\ref{Ai})
and (\ref{Ai2}) in eqs. (\ref{lageffsep1}), (\ref{lageffsep2}). We
get
\begin{equation}
\label{pretri}
\begin{split}
\mathcal{L}_{eff}^{3} & = - i g \left\{ \left[\tilde{W}_{\mu \nu}^{3} \tilde{W}_{}^{+ \, \mu} \tilde{W}_{}^{- \, \nu} + \tilde{W}_\mu^3 (\tilde{W}_{}^{- \, \mu \nu} \tilde{W}_{\nu}^{+} - h.c.) \right] \right. \\
- & \frac{g^2}{\overline{G}^2 \overline{M}^2} \left[ (\square_3 + \square_+ + \square_-)(\tilde{W}_{\mu \nu}^{3} \tilde{W}_{}^{+ \, \mu} \tilde{W}_{}^{- \, \nu} + \tilde{W}_{\mu}^{3}(\tilde{W}_{}^{- \mu \nu} \tilde{W}_{\nu}^{+} - h.c.))\right. \\
- & \partial_\mu \tilde{W}_{\nu}^{3} (\partial^\mu \partial_\rho \tilde{W}_{}^{+ \, \rho} \tilde{W}_{}^{- \, \nu} - h.c.) + \partial_\mu \partial_\rho \tilde{W}_{}^{3 \, \rho} (\partial^\mu \tilde{W}_{\nu}^{+} \tilde{W}_{}^{- \, \nu} - h.c.)\\
- & \tilde{W}_\mu^3 (\partial_\nu \tilde{W}_{}^{+ \, \mu} \partial^\nu \partial_\rho \tilde{W}_{}^{- \, \rho} - h.c.) + \partial_\mu \tilde{W}_{\nu}^{3} (\partial^\nu \partial_\rho \tilde{W}_{}^{+ \, \rho} \tilde{W}_{}^{- \, \mu} - h.c.) \\
- & \left. \left. \partial_\mu \partial_\rho \tilde{W}_{}^{3 \,
\rho} (\partial_\nu \tilde{W}_{}^{+ \, \mu} \tilde{W}_{}^{- \, \nu}
-h.c.) + \tilde{W}_\mu^3 (\partial^\mu \tilde{W}_{\nu}^+
\partial^\nu \partial_\rho \tilde{W}_{}^{- \, \rho} - h.c.) \right]
\right\},
\end{split}
\end{equation}
where $\square_+, \, \square_-, \, \square_3$ operate only
 on the fields which bear the same index.

In order to make explicit the new physics anomalous contributions to
the trilinear couplings, we will rewrite eq. (\ref{pretri}) in terms
of the mass eigenstates. First of all, we introduce $\tilde{A}$ and
$\tilde{Z}$ fields from $\tilde{W}^3$ and $\tilde{Y}$ in the usual
way:
\begin{equation}
\label{basischange} \binom{\tilde{W}^3}{\tilde{Y}} =
\genfrac{(}{.}{0pt}{}{c_{\theta}}{-s_{\theta}}
\genfrac{.}{)}{0pt}{}{s_{\theta}}{\  c_{\theta}}
\binom{\tilde{Z}}{\tilde{A}}.
\end{equation}
Substituting in eq. (\ref{eq:3}) we get
\begin{equation}
\begin{split}
\label{eq:4}
\mathcal{L}_{eff}^2  & = - \frac{1}{2} \tilde{W}_{\mu \nu}^+ \tilde{W}^{- \, \mu\nu} - \frac{1}{4} \tilde{Z}_{\mu \nu} \tilde{Z}^{\mu \nu} - \frac{1}{4} \tilde{A}_{\mu \nu} \tilde{A}^{\mu \nu} \\
+ & \tilde{M}_W^2 \tilde{W}_\mu^+ \tilde{W}^{- \, \mu} + \frac{\tilde{M}_Z^2}{2}\tilde{Z}_\mu \tilde{Z}^\mu+\frac{1}{2\overline{M}^2}[z_w \tilde{W}_{\mu \nu}^+ \square \tilde{W}^{- \, \mu \nu}\\
+ & \frac{z_\gamma}{2} \tilde{A}_{\mu \nu} \square
\tilde{A}^{\mu\nu} + \frac{z_z}{2} \tilde{Z}_{\mu \nu} \square
\tilde{Z}^{\mu \nu} + z_{z \gamma} \tilde{A}_{\mu \nu} \square
\tilde{Z}^{\mu \nu}]
\end{split}
\end{equation}
where
\begin{equation}
\begin{split}
& z_w = \frac{g^2}{\overline{G}^2}, \ z_\gamma = \frac{2 e^2}{\overline{G}^2}, \quad \textrm{with} \quad e = g s_{\theta} = g' c_{\theta},\\
z_z = & \frac{g^2 (c_{\theta}^4 + s_{\theta}^4)}{c_{\theta}^2
\overline{G}^2}, \ z_{z \gamma} = \frac{g g' c_{2
\theta}}{\overline{G}^2}, \ \tilde{M}_W^2 = \frac{v^2 g^2}{4}, \
\tilde{M}_Z^2 = \frac{v^2(g^2+g'^{2})}{4}. \label{z}
\end{split}
\end{equation}
We then introduce the field rescaling:
\begin{gather}
\tilde{W}_\mu^{\pm} = \left(1+ \frac{z_w}{2}(\frac{\square}{\overline{M}^2} - \frac{\tilde{M}_W^2}{{\overline{M}^2}}) \right)W_\mu^\pm,  \notag \\
\tilde{Z}_\mu = \left(1 + \frac{z_z}{2}(\frac{\square}{\overline{M}^2} - \frac{\tilde{M}_Z^2}{\overline{M}^2}) \right)Z_\mu, \label{rin}\\
\tilde{A}_\mu = \left(1 +
\frac{z_\gamma}{2}\frac{\square}{\overline{M}^2} \right)A_\mu + z_{z
\gamma} \frac{\square}{\overline{M}^2} Z_\mu, \notag
\end{gather}
which allows us to get rid of the anomalous ``$\square$'' terms in
the quadratic part of the Lagrangian. We then obtain
\begin{equation}
\mathcal{L}_{eff}^2 = -\frac{1}{2} W_{\mu \nu}^+ W^{- \, \mu\nu} -
\frac{1}{4} Z_{\mu \nu} Z^{\mu \nu} - \frac{1}{4} A_{\mu \nu} A^{\mu
\nu} + M_W^2 W_\mu^+ W^{- \, \mu} + \frac{M_Z^2}{2} Z_\mu Z^\mu,
\end{equation}
where:
\begin{equation}
M_{W}^2 = \tilde{M}_W^2 \left(1 - z_w
\frac{\tilde{M}_W^2}{\overline{M}^2} \right),\quad M_{Z}^2 =
\tilde{M}_Z^2 \left(1 - z_z \frac{\tilde{M}_Z^2}{\overline{M}^2}
\right).
\end{equation}
This is just the SM electroweak gauge boson bilinear Lagrangian;
however, the rescaling (\ref{rin}) of the fields will affect both
the couplings with fermions and the trilinear bosonic couplings.
Let's shift to the $\tilde{W}^\pm, \ \tilde{A}, \ \tilde{Z}$ basis
in eq. (\ref{ferm}), then rescale the fields according to
(\ref{rin}). We get:
\begin{equation}
\begin{split}
& \mathcal{L}_{charged} = - \frac{e}{\sqrt{2} s_{\theta}}
\overline{\psi}_u \gamma^\mu (1-\frac{\gamma^5}{2})
\left(1 + \frac{z_w}{2}(\frac{\square}{\overline{M}^2} - \frac{\tilde{M}_W^2}{\overline{M}^2})\right) \psi_d W_\mu^+ + \, h.c.\\
& \mathcal{L}_{neutral} =  - \frac{e}{s_{\theta} c_{\theta}}
\left(1+\frac{z_z}{2}(\frac{\square_Z}{\overline{M}^2} -
\frac{\tilde{M}_Z^2}{\overline{M}^2}) \right) \overline{\psi}
\gamma^\mu \Big[\frac{\tau^3}{2} \frac{(1-\gamma^5)}{2}
\\
& \qquad \qquad ~- Q s_{\theta}^2 \left(1+
\frac{c_{\theta}}{s_{\theta}} z_{z
\gamma}\frac{\square_Z}{\overline{M}^2} \right) \Big] \psi  Z_\mu -
e \overline{\psi} \gamma^\mu Q \psi \left(1 -
\frac{z_\gamma}{2}\frac{\square}{\overline{M}^2} \right) A_\mu,
\label{Leff}
\end{split}
\end{equation}
where again we use the convention that $\square_Z$ does only operate on $Z$ and $Q = \frac{\tau^3}{2} + \frac{B-L}{2}$.\\
We see that the photon-fermion interaction at zero momentum
correctly predicts $e$ as the physical value of the electric charge.
The Fermi constant $G_F$ can be measured from the $\mu$ decay, still
at zero momentum. We have:
\begin{equation}
\begin{split}
\label{GF}
\frac{G_F}{\sqrt{2}} & = \frac{e^2}{8s_{\theta}^2} \left(1-z_w
\frac{M_W^2}{\overline{M}^2}\right) \frac{1}{{\tilde M}_W^2} \left(1 + z_w \frac{M_W^2}{\overline{M}^2} \right)\\
= & \frac{e^2}{8s_{\theta}^2 c_{\theta}^2 M_Z^2} \left(1+ z_z
\frac{M_Z^2}{\overline{M}^2}\right),
\end{split}
\end{equation}
where we have substituted the physical masses $M_W$ and $M_Z$ to
$\tilde{M}_W$ and $\tilde{M}_Z$ since they only differ by terms of
$O(M_Z^2/\overline{M}^2)$, which are negligible in a term which is
already of the same order. From eq. (\ref{GF}) we can define the
effective Weinberg angle (see \cite{Altarelli:1991fk}):
\begin{equation}
\frac{G_F}{\sqrt{2}} = \frac{e^2}{8s_{\theta_0}^2 c_{\theta_0}^2
M_Z^2} \ \Rightarrow \ s_{\theta_0}^2 c_{\theta_0}^2 = s_{\theta}^2
c_{\theta}^2 \left(1 + z_z \frac{M_Z^2}{\overline{M}^2} \right),
\label{sctheta}
\end{equation}
that is
\begin{equation}
 s_{\theta_0}^2 = s_{\theta}^2 \left(1 + \frac{c_{\theta}^2}{c_{2 {\theta}}}z_z \frac{M_Z^2}{\overline{M}^2} \right). \label{stheta}
\end{equation}

Notice that the $\epsilon$ parameters, which we obtained from the
correlators (\ref{corr}), can be also derived as in
\cite{Casalbuoni:1995qt} by using the rescaling of the fields given
in  (\ref{rin}) and by evaluating  the $\Delta \rho$,
$\Delta k$ and $\Delta r_W$ parameters  \cite{Altarelli:1991fk}. The
results precisely agree with those obtained in Section
\ref{universal}.

Now, to get the corrections to the trilinear gauge boson couplings,
it is sufficient to substitute eq. (\ref{rin}) in eq.
(\ref{pretri}). The general expression we get is quite long, so we
will not report it. We will instead specialize  to the study of a
particular physical process, which will allow us to simplify eq.
(\ref{pretri}) slightly, in order to get a more readable result. The
process we will consider as an example is $e^+e^- \rightarrow
W^+W^-$ scattering; in this case, the $W$ are on-shell, so that we
have
\begin{equation}
\partial_\mu W^{\pm \, \mu} = 0
\end{equation}
thanks to the Ward identity. Limiting the study to tree level, we
can either have a virtual $\gamma$ or a virtual $Z$ as an
intermediate state (besides the neutrino  exchange which is not
relevant to the study of the trilinear gauge couplings). The
4-divergence of the virtual $\gamma$ also vanishes due to the Ward
identity; while the virtual $Z$ has an approximately vanishing
transverse contribution thanks to the Dirac equation, since it is
coupled to external fermions of negligible mass (compared to the
center of mass energy which is of order $M_Z$). So we can take:
\begin{equation}
\partial_\mu A^\mu = 0, \quad \partial_\mu Z^\mu \simeq 0.
\end{equation}
As a consequence all the divergence-proportional terms in the
effective Lagrangian (the last three lines in eq. (\ref{pretri}))
can be safely dropped out. In this way, taking into account eq.
(\ref{stheta}), we get the following expression for the trilinear
gauge boson couplings, relevant for the $e^+e^- \to W^+ W^-$
process:
\begin{equation}
\begin{split}
\label{finale}
\mathcal{L}_{eff}^3 & = -ie \frac{c_{\theta_0}}{s_{\theta_0}}\left(1 + \frac{z_z}{2c_{2\theta}} \frac{M_Z^2}{\overline{M}^2} - \frac{z_w}{2} \frac{\square_+ + M_W^2}{\overline{M}^2} - \frac{z_w}{2}   \frac {\square_- + M_W^2}{\overline{M}^2} \right. \\
- & \left. \frac{z_z}{2}\frac{\square_Z + M_Z^2}{\overline{M}^2} \right) \left(Z_{\mu \nu} W_{}^{- \, \mu} W_{}^{+ \, \nu} + Z_\nu(W_{}^{- \, \mu \nu} W_{\mu}^{+} - W_{}^{+ \mu \nu} W_{\mu}^{-}) \right) \\
+ & ie \left(1 - \frac{z_w}{2} \frac{\square_+ + M_W^2}{\overline{M}^2} -\frac{z_w}{2}\frac{\square_- + M_W^2}{\overline{M}^2} + (\frac{z_\gamma}{2} - z_w) \frac{\square_A}{\overline{M}       ^2}\right) \\
& \left(A_{\mu \nu} W^{- \, \mu} W^{+ \, \nu} + A_\nu(W^{- \mu \nu}
W_{\mu}^{+} - W_{}^{+ \mu \nu} W_{\mu}^{-}) \right).
\end{split}
\end{equation}
We see that the tensor structure of the anomalous terms is identical
to that of the SM, while the coefficients of the various operators
contain derivative terms. Due to the presence of these nontrivial
form factors and to the fact that the fermion-gauge boson couplings
are also modified, as shown in eq. (\ref{Leff}), the comparison of
the predictions of eq. (\ref{finale}) to the experimental data is
not direct, but  requires a full calculation of the $e^+ e^-
\to W^+ W^-$ cross-section in the GD-BESS model.
 However the present experimental bounds from LEP2 on the anomalous trilinear couplings
\cite{Yao:2006px} have errors of the order of a few percent. Since
the determination of the new physics parameters entering in eq. (\ref{finale})
is at the level of a few permil from LEP/Tevatron, it is clear that
a higher precision will be necessary in order to achieve the same accuracy.
We have nevertheless checked that, taking for example the expression
of $g_1^Z$ extracted from eq. (\ref{finale}), and  comparing with the
present experimental value for $g_1^Z=0.984^{+0.022}_{-0.019}$, we get 
 bounds
on the  plane $({\overline M},\frac{1}{\displaystyle \overline G})$ which
are not relevant with respect to the ones shown in Fig. \ref{fig:4}.
We have also checked that, in order to have comparable bounds, one would need
an extimation of $g_1^Z$ at the permil level.


\section{Conclusions}
\label{conclusion}

We have considered a linear moose model based on the extended gauge
symmetry $SU(2)^{2N+1}$. The model has the central link missing and
a left-right symmetry along the moose. As a consequence of the
missing link the $\epsilon_3$ parameter is zero at the leading order
in $(M_Z^2/\overline{M}^2)$, where $\overline{M}$ is the mass scale
of the new resonances. This result can also be understood in the
following way: since the model describes $N$ pairs of new gauge
boson triplets degenerate in mass, the $\epsilon_3$ contribution of
the vector resonances is canceled by the axial vector one.

We have computed the low energy effective Lagrangian by eliminating
the internal moose gauge fields and extracted the electroweak
precision parameters and the trilinear anomalous couplings. Since
these parameters turn out to be of order
$(g/\overline{G})^2M_Z^2/\overline{M}^2$ and since the new effective
coupling $\overline{G}$ could well be much larger than $g$,  the
possibility of a low scale $\overline{M}$ is left open.
We expect the GD-BESS model in the present formulation
to become a strongly interacting
theory at energies of the order of 1.7 TeV independently of the values of the model parameters as a consequence of the perturbative unitarity violation,
so
it is interesting, among the future developments, to study how to unitarize it.
We are currently investigating the  possibility  to include  scalars
associated to the non linear $\sigma$-model fields. However,  in this unitarized extension,  we
expect corrections to the $\epsilon_i$ parameters of the same order
of magnitude as the ones evaluated here so not spoiling our overall
conclusions.


\end{document}